\documentclass[english]{paper}
\usepackage{ae,aecompl}
\usepackage{helvet}
\usepackage[T1]{fontenc}
\usepackage[latin9]{inputenc}
\usepackage{amsmath}
\usepackage{amssymb}
\usepackage{graphicx}
\usepackage{esint}

\makeatletter
\usepackage{float}

\usepackage{babel}

\usepackage{babel}

\usepackage{babel}

\makeatother

\usepackage{babel}
\begin{document}

\title{Level statistics of disordered spin-$\frac{1}{2}$ systems and its
implications for materials with localized Cooper pairs. }

\author{Emilio Cuevas$^{1}$ \& Mikhail Feigel'man$^{2,3}$ \& Lev Ioffe$^{4}$
\& Marc Mezard$^{5}$}

\maketitle
$^{1}$Departamento de Física, Universidad de Murcia, E-30071 Murcia,
Spain

$^{2}$L. D. Landau Institute for Theoretical Physics, Kosygin str.2,
Moscow 119334, Russia

$^{3}$Moscow Institute of Physics and Technology, Moscow 141700,
Russia

$^{4}$Department of Physics and Astronomy, Rutgers University, 136
Frelinghuysen Rd., Piscataway, NJ 08854, USA

$^{5}$CNRS, Université Paris-Sud, UMR 8626, LPTMS, Orsay Cedex, F-91405
France 
\begin{abstract}
\textbf{The origin of continuous energy spectra in large disordered
interacting quantum systems is one of the key unsolved problems in
quantum physics. While small quantum systems with discrete energy
levels are noiseless and stay coherent forever in the absence of any
coupling to external world, most large-scale quantum systems are able
to produce thermal bath and excitation decay. This intrinsic decoherence
is manifested by a broadening of energy levels which aquire a finite
width. The important question is what is the driving force and the
mechanism of transition(s) between two different types of many-body
systems - with and without intrinsic decoherence? Here we address
this question via the numerical study of energy level statistics of
a system of spins-$\frac{1}{2}$ with anisotropic exchange interactions
and random transverse fields. Our results present the first evidence
for a well-defined quantum phase transition between domains of discrete
and continous many-body spectra in a class of random spin models.
Because this model also describes the physics of the superconductor-insulator
transition in disordered superconductors like InO and similar materials,
our results imply the appearance of novel insulating phases in the
vicinity of this transition. } 
\end{abstract}
Although quantum phase transitions between regimes with and without
intrinsic decoherence were first discussed by Anderson in 1958~\cite{Anderson1958}
, their understanding is still incomplete. More recently, a number
of works have studied these transitions in strongly disordered electron
systems with weak repulsion\cite{Altshuler1997,Gornyi2005,Basko2006}
and reached the conclusion\cite{Gornyi2005,Basko2006} that a finite-temperature
phase transition takes place between two regimes, {}``weakly insulating''
and {}``strongly insulating'', that are characterized by non-zero
and zero conductivities respectively. In terms of the many-particle
excitation spectra this result translates into the existence of an
\textit{extensive} energy threshold, with a critical excitation energy
$\mathcal{E}_{c}\propto\mathcal{V}T_{c}$ that separates the discrete
spectrum of excitations at $E<\mathcal{E}_{c}$ and continous one
at $E>\mathcal{E}_{c}$ for any large but finite subsystem with $\mathcal{V}$
degrees of freedom. It also implies that the low-temperature state
is free from decoherence because the lifetime of all excitations is
infinite. The phase transition predicted in the work~\cite{Basko2006}
was not observed yet; instead, in hopping insulators one usually observes
that conductivity vanishes continously at $T\rightarrow0$: $\sigma(T)\sim e^{-(T^{*}/T)^{a}}$,
with $a<1$.

Different physical properties of the excitations at low and high energies
in the infinite system are reflected in different statististical properties
of the spectra of finite systems at $E<\mathcal{E}_{c}$ and at $E>\mathcal{E}_{c}$.
Intuitively, if the eigenvectors are extended, as expected for the
state where local excitations decay, they are subject to inter-level
repulsion. Conversely, if the eigenvectors are localized, eigenvalues
corresponding to excitations localized in different parts of the system
are independent and one expects a Poisson distribution of energy levels.

Let us give a general qualitative argument in supporting this statement.
Consider a small perturbation of the Hamiltonian that controls the
dynamics of a generic quantum system in thermodynamic limit: 
\begin{equation}
H\rightarrow H(1+\dot{\phi}(t,x))\label{eq:H_variation}
\end{equation}
 where $\phi(t,x)$ is a generic slow function of coordinates and
time. A small perturbation of this type results in the slow (adiabatic)
motion of energy levels $E_{n}(t)$. In the absence of level repulsion,
different levels cross without affecting each other, so that this
motion leads only to the total phase of the wave function. Because
the field $\dot{\phi}(t,x)$ (which is similar to the gravitational
potential\cite{Luttinger1964}) is conjugated to the energy density,
the absence of response to it implies absence of the energy flux.
An excitation with energy $\Delta E$ localized around point $x$
acquires phase $\exp(-i\Delta E\phi(t,x))$ due to perturbation (\ref{eq:H_variation});
in contrast, a delocalized excitation becomes a superposition of other
excitations. Thus, the absence of the response also implies that excitations
are localized and do not decay. We conclude that the absence of level
repulsion implies the localization of excitations, absence of their
decay and of the energy flux, i.e. formation of a strong insulator.
Because level statistics can be studied for relatively small systems,
this correspondence between level statistics and physical properties
in thermodynic limit provide a convenient numerical tool to predict
the properties of physical systems. This strategy has been used for
instance in \cite{Oganesyan2007}.

One of the best experimental systems to test these general ideas is
provided by strongly disordered superconductors (InO, TiN) in the
vicinity of the disorder-induced superconductor-insulator transition~\cite{Shahar1992};
for recent reviews see~\cite{Gantmakher2010} and the introduction
of the paper~\cite{Feigelman2010a}. One advantage of these systems
is that the transition can be fine-tuned by a magnetic field. In the
insulating state they demonstrate purely activated behaviour $\sigma(T)\sim e^{-(T^{*}/T)}$,
with $a=1$.\cite{Sambandamurthy2004,Ovadia2009} Far in the insulating
state~\cite{Shahar1992} this behaviour can be understood in terms
of a single electron pseudogap~\cite{Sacepe2011a,Sacepe2010} that
results from binding of localized electrons~\cite{Feigelman2007,Feigelman2010a}.
However, the persistance of an activated behavior in the vicinity
of the superconductor-insulator transition (SIT) where one expects
the appearance of low energy Cooper pairs requires another explanation.
This was proposed in the work~\cite{Ioffe2010,Feigelman2010b} in
terms of a collective mobility edge $\epsilon_{c}$ which separates
the domain of localized excitations with energies $\omega<\epsilon_{c}$
from the domain of delocalized modes with $\omega>\epsilon_{c}$,
which serves as an intrinsic thermal bath. The threshold $\epsilon_{c}(g)$
depends on the parameter characterizing disorder, $g$. It vanishes
at $g=g_{c}$ at which superconductivity appears. At $g<g_{c}$ the
threshold energy $\epsilon_{c}$ is \textit{intensive} \cite{Ioffe2010,Feigelman2010b}
(it does not grow with the size of the system). In this regime, the
density of excitations is proportional to $\exp(-\epsilon_{c}/T)$,
leading to the activated behaviour of conductivity. At even stronger
disorder, $g=g^{*}$, the threshold energy $\epsilon_{c}(g)$ diverges.

Logically there are two possibilities; the first one is that the divergence
of the threshold $\epsilon_{c}$ at $g\rightarrow g^{*}$ is cut off
by the system size, so that at stronger disorder $\epsilon_{c}$ scales
with the system volume, $\mathcal{V}$. In this situation one expects
the finite temperature transition between insulator and hard insulator
predicted in \cite{Gornyi2005,Basko2006}. In the second scenario,
the energy scale $\epsilon_{c}$ becomes really infinite at $g\rightarrow g^{*}$,
which implies a complete localization and the absence of a thermal
equilibrium at all temperatures. We will show below that both scenarios
can be realized in discrete spin models depending on the detailed
form of the Hamiltonian: in the models with purely local density-density
interaction $\epsilon_{c}$ is infinite at $g<g^{*}$, but the addition
of non-local density-density interaction terms results in the appearance
of a narrow range $g_{I}<g<g^{*}$ where $\epsilon_{c}$ is extensive,
signaling the finite temperature transition in this regime. This result
raises the possibility that a finite temperature transition between
strong and weak insulators might be observed in InO, TiN or similar
films very close to the superconductor-insulator transition.

The theoretical work \cite{Ioffe2010,Feigelman2010b} is based on
several approximations. First, it assumes the presence of a strong
pseudogap, which allows one to reduce the Hilbert space of the full
electron problem to the one spanned by \textquotedbl{}pseudospin\textquotedbl{}-$\frac{1}{2}$
variables $s_{j}^{\pm},s_{j}^{z}$, that describe~\cite{Anderson1959}
creation-annihilation and counting operators of the localized Cooper
pairs. This assumption is borne out by experimental results\cite{Sacepe2011a}.
On the theoretical side, its origin lies in the fractal stucture of
the wave functions near to the mobility edge of the single electron
problem\cite{Feigelman2007,Feigelman2010a}. It is further supported
by direct numerical simulations of electrons with attractive interactions
in a random potential\cite{Bouadim2011}. The second approximation
is the use of a recursion-equation technique, which becomes exact
only for a special tree-like structure of a lattice completely devoid
of small loops (Bethe lattice). Finally, the formalism developed in
these works does not include explicitely collective excitations of
high energy involving many spins. Therefore this formalism does not
allow to completely settle whether the mobility edge is intensive
or extensive. Note that although another prediction of our formalism,
namely the anomalous broadening of the superconducting order parameter
distribution in the vicinity of transition, was confirmed experimentally
\cite{Sacepe2011a}, no such data exist for insulating phase.

\begin{figure}
\includegraphics[width=3in]{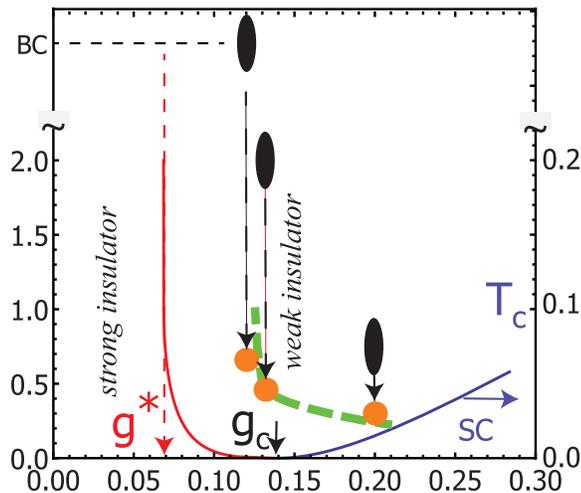}

\caption{Phase diagram of a strongly disordered superconductor with a large
pseudogap as a function of interaction constant $g$. The full lines
show the predictions of the analytical theory of the model (\ref{H_A})
for the critical temperature (right vertical axis) and the threshold
energy, $\epsilon$, (left axis) of spin flip excitations in infinite
random graph with $Z=3$ neighbors. The vertical ovals show the values
of the critical coupling constant that correspond to a transition
between different types of spectra for different energies $E$ in
the finite random graph model of a small size ($N=16-20)$ as determined
by direct numerical simulations. The uppermost oval shows the transition
at the many-body band center (corresponding to $E\gg1$) that sets
a lower bound for the critical $g(E)$. The thick dashed line shows
the position of the spectral threshold for single-spin excitations
with energy $\epsilon$ adjusted by finite-size effects, as explained
in the main text and in the Supplement \ref{sub:Finite-size-effect-upon}.
The small circles show the typical energy of the single-spin excitations,
$\epsilon(E)$, that give the main contribution to the many body excitations
studied in direct numerical simulations. The good agreement between
their position and expectations (dashed line) confirms the validity
of the cavity method~\cite{Ioffe2010,Feigelman2010b} that is used
to obtain the results in large systems. The very small change in the
critical value of the coupling constant between excitations at energy
$E\approx2.0$ and the band center implies that all excitations, at
high and low energies, become localized when $g<g^{*}$.}

\label{Flo:PhaseDiagram} 
\end{figure}

\begin{figure}
\includegraphics[width=3in]{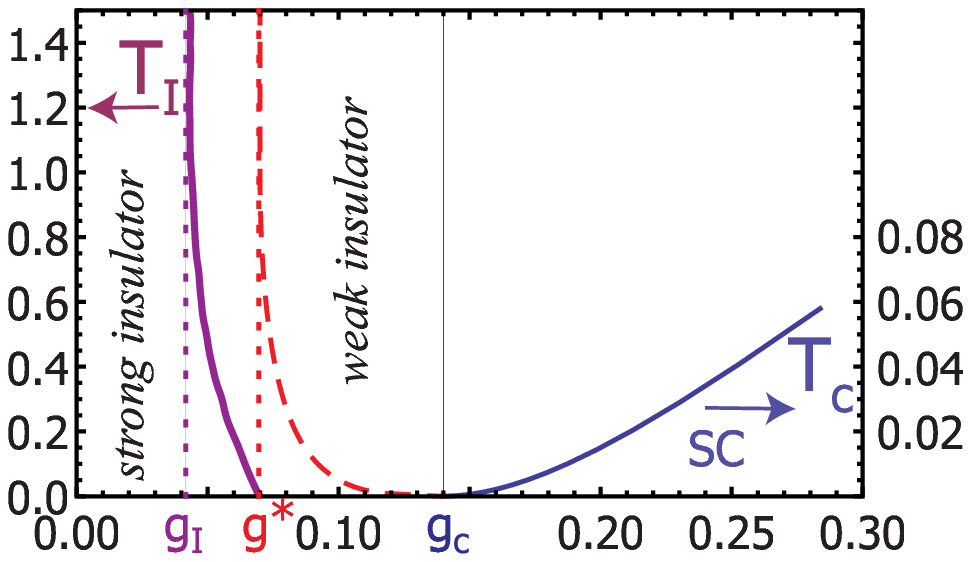}

\caption{Phase diagram in the temperature - coupling constant plane for the
model (\ref{H_AA}) with $Z=3$ ($K=2$), obtained from the solution
of cavity equations. The strength of the $s^{z}s^{z}$ interaction
is $J^{z}=0.1$. This interaction leads to a qualitatively new phenomenon,
the appearance of a finite temperature transition between weak and
strong insulators. In the weak insulator, excitations at sufficiently
high energies can decay even at zero temperature. A non-zero temperature
results in non-zero relaxation of all excitations, even the ones of
lowest energy. In contrast, in the strong insulator, no excitation
with intensive energy can decay. As the interaction constant is decreased,
the temperature separating these phases goes to infinity at $g=g_{I}.$
At smaller coupling $g<g_{I}$, all excitations, even those with \textit{extensive}
energy remain localized. The value of $g_{I}\approx0.042$ is approximately
equal to $0.30g_{c}$. The ratio $g_{I}/g_{c}=0.3$ is in good agreement
with the results of the direct diagonalization on small graphs, as
can be seen from the critical values of $J_{c}^{XY}$ found in Fig.\ref{Flo:r(J)}:
$0.02$ for the middle of the band band and $0.74$ for the low energies. }

\label{Flo:PhaseDiagramSzSz} 
\end{figure}

In the present Letter we provide a direct numerical proof of the validity
of the many-body localization scenario developed in\cite{Ioffe2010,Feigelman2010b}
for a more realistic random spin lattice and taking into account the
whole excitation spectrum. Our basic model contains spins $\frac{1}{2}$
that describe the absence or presence of Cooper pair on a localized
single electron state. These spins are subject to random fields along
the $z$ direction and are coupled by XY interactions (\ref{H_A}).
The main physical result of our study are the phase diagrams, shown
in Figs. \ref{Flo:PhaseDiagram} and \ref{Flo:PhaseDiagramSzSz},
in terms of the dimensionless transverse spin coupling $g$ and excitation
energy $\omega$ (Fig.\ref{Flo:PhaseDiagram}) and temperature (Fig.\ref{Flo:PhaseDiagramSzSz}).
A \textit{size-independent (non-extensive)} threshold energy $\epsilon_{c}(g)$
is found in this \textquotedbl{}minimal\textquotedbl{} model, Eq.(\ref{H_A}),
for $g\in(g^{*},g_{c})$, while all eigenstates are localized at $g<g^{*}$.
This implies that, for this Hamiltonian, the transition line separating
the weak and hard insulators does not depend on temperature. On the
other hand, we shall show that, in a model where the Hamiltonian is
modified and includes additional interactions between the $z$ components
of spins, the threshold energy remains finite but \emph{extensive}
in finite systems in some range of $g<g^{*}$, see Fig. \ref{Flo:PhaseDiagramSzSz}.
This leads to the temperature driven transition similar to the one
predicted in ~\cite{Gornyi2005,Basko2006}.

The phase diagrams shown in Figs. \ref{Flo:PhaseDiagram} and \ref{Flo:PhaseDiagramSzSz}
have another important feature: they predict a direct transition between
superconductor and insulator with characteristic energy scales that
go down to zero continously on both sides of the transition, similar
(but different in the important details) to the transition expected
in dual theories\cite{Fisher1990a,Fisher1990b}. In the superconducting
state this energy scale is given by the typical value of the order
parameter or the transition temperature, while in the insulating states
it is the value of the threshold energy $\epsilon_{c}(g)$, which
implies the Arrhenius behavior of the resistivity at very low temperatures.
These predictions are in agreement with the detailed studies of the
transition driven by a magnetic field in films\cite{Sambandamurthy2005}
and in Josephson junction arrays with moderate $E_{J}/E_{c}$\cite{Fazio2001,Paramanandam2011}.

As discussed in details elsewhere\cite{Feigelman2010a,Feigelman2010b}
a superconductor with a large pseudogap and a weak long-range Coulomb
repulsion is faithfully described by the model of spins-$\frac{1}{2}$
: 
\begin{equation}
H=-W\sum_{i}\xi_{i}s_{i}^{z}-\sum_{(ij)}J_{ij}^{xy}(s_{i}^{+}s_{j}^{-}+s_{i}^{-}s_{j}^{+})\label{H_A}
\end{equation}
 where points $i,j$ belong to a random graph $G$ with a fixed coordination
number $Z$, $\mathbf{s}_{i}=\frac{1}{2}\mathbf{\sigma}_{i}$ are
spin-$\frac{1}{2}$ operators, $s_{i}^{\pm}=s_{i}^{x}\pm is_{i}^{y}$,
the sum $\sum_{(ij)}$ goes over all different pairs of nearest neighbors
$i,j$ on $G$, and all nonzero matrix elements are equal to $J_{ij}^{xy}=Wg/(Z-1)$.
The random energies $\xi_{j}$ are uncorrelated at different sites
and choosen from the box distribution $Q(\xi)=\frac{1}{2}\theta(1-|\xi|)$,
corresponding to bandwidth $W$.

Eigenstates of the Hamiltonian (\ref{H_A}) are vectors in the $2^{N}$-dimensional
Fock space ($N$ is the total number of sites of $G$). As discussed
above, the transition between decoherent and coherent states can be
deduced from the change in the statistics of the exact eigen-energies
$E_{i}$ of the Hamitlonian. To identify the energy level statistics,
we study the dimensionless parameter $r_{n}\in[0,1]$ defined as 
\begin{equation}
r_{n}=\min(\delta_{n},\delta_{n-1})/\max(\delta_{n},\delta_{n-1})\label{rn}
\end{equation}
 where $\delta_{n}=E_{n}-E_{n-1}$ and $E_{n}$ is the $n$-th energy
level. The average value $r=\langle r_{n}\rangle$ is equal to $0.38$
for Poisson level statistics, and to $0.53$ for Wigner-Dyson statistics.
In the limit of infinite systems one expects a sharp transition between
these two values as a function of $g$, or of the energy. In a finite
system the parameter $r$ increases smoothly, but the curve $r(g)$
is expected to become steeper as the size is increased.

\begin{figure}
\includegraphics[width=4in]{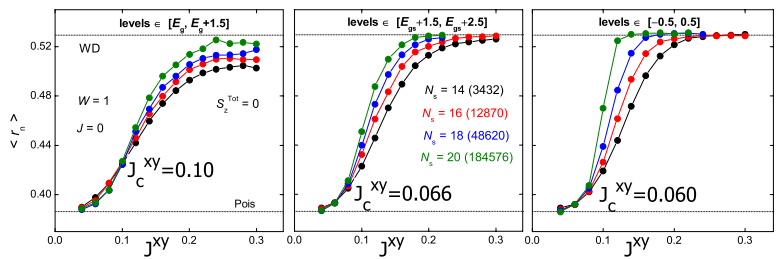}

\caption{The energy-level statistics is characterized by the average $\left\langle r_{n}\right\rangle $
that distinguishes Wigner-Dyson and Poisson distributions (values
of $\left\langle r_{n}\right\rangle $ corresponding to these distributions
are shown by dashed lines). The left panel shows the statistics of
the low-energy excitations in the energy interval $(E_{gs},E_{gs}+1.5)$
as a function of the transverse interaction constant, $J^{xy}$, for
the $Z=3$ random graph with bandwidth $W=1$. The middle panel shows
similar results for intermediate energies, and the right panel corresponds
to high energies, close to the center of the many-body spectrum. }

\label{Flo:r(J)} 
\end{figure}

The appearance of a unique crossing point of these curves for different
sizes $N$ implies a well-defined phase transition in the limit of
infinite-size systems. This is seen in the exact diagonalization of
the Hamiltonian (\ref{H_A}) with $Z=3$, shown in Fig. \ref{Flo:r(J)}.
Apart from the persistance of the transition for small system sizes,
these data also show that the critical value of $J^{xy}$ changes
very little as the energy is increased to the center of the band.
This proves that high energy states in this model become localized
together with the low energy ones, in more physical terms it implies
that no new decay channels appear at high energies. The absence of
new decay processes at high energies is the physical reason why in
this model one does not get the intermediate phase with \emph{extensive}
energy threshold. Also, these data show that the transition happens
at a value $r_{n}(g)$ that is close to its value $0.38$ expected
for Poisson statistics. This is due to a large distribution of relaxation
rates in these systems\cite{Feigelman2010b} that implies that in
case of small systems many realizations of the random energies give
localized states, while delocalization only happens with a small probability
as we explain in a more detail below.

In order to compare quantitatively the results of the direct diagonalization
shown in Fig. \ref{Flo:r(J)} with the predictions of the theory \cite{Feigelman2010b}
we need to take into account the finite size effects. They are very
significant for the sizes (number of spins $N\leq20$) available for
direct diagonalization, for two reasons. First, in a finite system
the crossover from Poisson to Wigner-Dyson statistics is expected
to occur when the level spacing $\delta(E)=\nu^{-1}(E)$ becomes approximately
equal to the level width $\Gamma(E)$ expected theoretically. The
latter quickly becomes exponentially small\cite{Feigelman2010b} when
the transition in the infinite system is approached, so that the condition
$\delta(E)=\Gamma(E)$ in a finite (not very large) system is satisfied
far away from the transition in the infinite system. Second, the level
widths fluctuate wildly from one finite system to another in small
systems. Because the infinite system can be viewed as being composed
of small ones, each of these parts having many neighbors, even a small
probability to find a delocalized (finite width) level in a small
system is sufficient for delocalization in the infinite system. This
makes the typical decay rate in the infinite system much larger than
the one in a collection of small ones. Both effects combine to push
up the apparent critical value of $g$ by roughly a factor of two.
Finally, for the quantitative comparison, we need to take into accout
that the critical point condition $\delta(E)=\Gamma(E)$ applies to
the level spacing and total level width in the many body systems,
whereas the analytical theory \cite{Feigelman2010b} gives the level
width of individual spin flips. The typical level at energies $E\gtrsim W$
consists of a few spin flips, which decay rates add to the total width
$\Gamma(E)$. The main contribution to the typical level at energy
$E\gtrsim W$ comes from single-spin excitations in a relatively narrow
energy window $\epsilon(E)\approx\sqrt{EW/N}$, see Supplement \ref{sub:Density-of-states}.
This allows us to interpret the crossing point in exact diagonalization
as the delocalization of individual spin flips as shown in Fig. \ref{Flo:PhaseDiagram}.
We present the details in the Supplement \ref{sub:Finite-size-effect-upon}. 

These results are consistent with the recent findings of paper~\cite{Scardiccio2011}
which studied a model similar to our (\ref{H_A}) but with identical
$s_{i}^{+}s_{j}^{-}$-couplings between all pairs of spins, allowing
exact integrability of the Hamiltonian. The conclusion reached in
this work is that many-body delocalized state disappears in exactly
integrable case but reappears when variations of couplings between
spins are allowed that destroys integrability. In the latter case
one expects a sharp transition between localized and delocalized regimes
similar to the one  found in model (\ref{H_A}).

We now discuss the generality of the conclusions reached above. The
absence of an intermediate phase with extensive energy threshold can
be traced back to the irrelevance of interactions between individual
spin flips. One can thus expect new physics to appear if such interaction
is introduced. The simplest and most physical model that has an additional
interaction between spin flips includes an additional \textquotedbl{}longitudinal\textquotedbl{}
spin-spin interaction: 
\begin{equation}
\tilde{H}=-W\sum_{i}\xi_{i}s_{i}^{z}-\sum_{(ij)}J^{z}s_{i}^{z}s_{j}^{z}-\sum_{(ij)}J_{ij}^{xy}(s_{i}^{+}s_{j}^{-}+s_{i}^{-}s_{j}^{+})\label{H_AA}
\end{equation}
 that can be viewed as due to the long range part of Coulomb interaction
($J^{z}<0$) or to phonon mediated attraction between electrons (($J^{z}>0$)
\cite{Feigelman2010a}. As will be clear below, the results do not
depend on the sign of $J^{z}\ll W$. The first two terms in the Hamiltonian
(\ref{H_AA}) describe a classical Ising magnet in a random field.
We shall be interested in its disordered phase which is realized when
the longitudinal interaction is small, $J^{z}\ll W$. In this case
the classical eigenstates of this magnet coincide with that of independent
spins, with energies that are weakly modified by the interaction.
In the absence of transverse interactions the excitations of this
magnet are individual spin flips with energies $\epsilon_{i}=W\xi_{i}+J^{z}\sum_{j(i)}s_{j}^{z}$.
In the ground state the direction of almost all spins is determined
by the sign of the random field, so the presence of a small $J^{z}$
affects weakly the density of states of these low-energy excitations
and their decay due to transverse interaction. Thus we expect that
the low-energy properties of the spectrum remain similar to the model
with zero $J^{z}$. The situation is very different at high energies
when the decay of a given spin occurs against the background of different
spin configurations. Physically, a spin excitation at energy $\epsilon$
has a much larger chance to propagate through a given site if the
energy of the spin flip at this site is close to $\epsilon$. The
energy of a spin flip is distributed with the probability density
$Q(\xi)$, where $\xi=\epsilon/W$, if all surrounding spins are locked
into a fixed configuration. In contrast, at extensive energies, the
surrounding spins acquire many different configurations. This increases
the probability that a given site is in resonance with the excitation.
This effect can be also described as being due to the interaction
between individual spin flip processes that allow new channels for
the decay of these excitations. This should result in the large suppression
of the critical value of the interaction constant at high energies.

\begin{figure}
\includegraphics[width=4in]{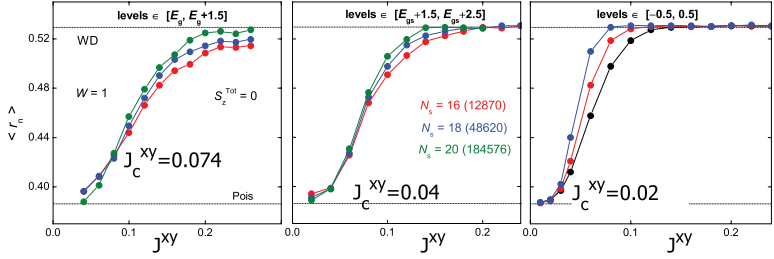}

\caption{Level statistics, characterized by the parameter $\left\langle r_{n}\right\rangle $
in the presence of weak longitudinal spin coupling $J$ for low energy
levels (left panel), intermediate energies (middle panel) and center
of the many body band (right panel). Even a small coupling $J^{z}=0.1$
has a significant effect, it shifts the transition to much smaller
values of the transverse coupling $g.$ }

\label{Flo:r_z(J)} 
\end{figure}

The exact diagonalization of the Hamiltonian on a $Z=3$ random graph
with $J^{z}=0.1$ confirms these expectations (Figure \ref{Flo:r_z(J)}).
In presence $J^{z}$, the value $g_{c}(0)$ is shifted slightly downwards
for low energies (($J_{c}=0.10\rightarrow0.074$). The shift becomes
substantial for medium energies ($J_{c}=0.066\rightarrow0.04$) and
very large for energies close to the band center: $J_{c}=0.06\rightarrow0.02$.
A slight shift of the critical value of the interaction constant at
low and medium energies is probably due to the fact that at these
energies a typical excited state contains more than one spin-flip
excitation. As explained above, the interaction between these excitations
leads to delocalization.

The solution of cavity equations, similar to those solved in~\cite{Feigelman2010b}
(see Supplement \ref{sub:SzSz-coupling-and}) confirms the appearance
of the extensive threshold for $J^{z}\neq0$ and gives the dependence
of the transition temperature $T_{I}(g)$ on the coupling constant
$g$ that we show in Fig. \ref{Flo:PhaseDiagramSzSz} for $J^{z}=0.1$.
At $g<g^{*}$ and $T<T_{I}(g)$ all excitations with intensive energies
are localized, no transport of any sort is possible. In contrast,
at higher temperatures $T>T_{I}(g)$ all excitations, even those with
low energies, acquire a non-zero width. At $g>g^{*}$ and zero temperature,
the excitations with low energy are localized while high energy excitations
decay. This latter distinction is smeared at any non-zero temperature
because the presence of even a small density of mobile high energy
excitations leads to a slow decay of even lowest energy ones.

The main conclusion of our work is the existence of two different
insulating phases and a prediction of the finite temperature transition
between a weak insulator characterized by an activated transport and
a strong one in which no transport occurs. Experimentally, the second
transition might be detected by the onset of an anomalously slow thermalization\cite{Canovi2011}.
These results are obtained within a lattice spin-$\frac{1}{2}$ model
which serves as a good approximation to the description of the superconductor-insulator
transition in a number of materials. We believe that it might be possible
to observe this transition in insulating films in a close proximity
to the superconducting transition. Due to a finite value of the pseudogap
in a realistic systems, a temperature-driven transition between weak
and strong insulating phases will be seen as a sharp crossover in
resistivity curves $R(T)$, with a rapid growth of the apparent activation
energy $d\ln R(T)/d(1/T)$ as $T$ decreases. Some preliminary experimental
evidence for this behaviour in TiN was reported in~\cite{Baturina2008a},
it was also observed in InO$_{x}$ ~\cite{Sacepe2011b}.

EC and MF thank the FEDER and Spanish DGI for financial support through
project FIS 2010-16430. MF acknowledges support through RFBR grant
10-02-00554 and RAS program \textquotedbl{}Quantum physics of condensed
matter\textquotedbl{}, LI was supported by ARO W911NF-09-1-0395, DARPA
HR0011-09-1- 0009 and NIRT ECS-0608842.

\appendix

\section{Supplementary online material}

\subsection{Density of states\label{sub:Density-of-states}}

\begin{figure}
\includegraphics[width=4in]{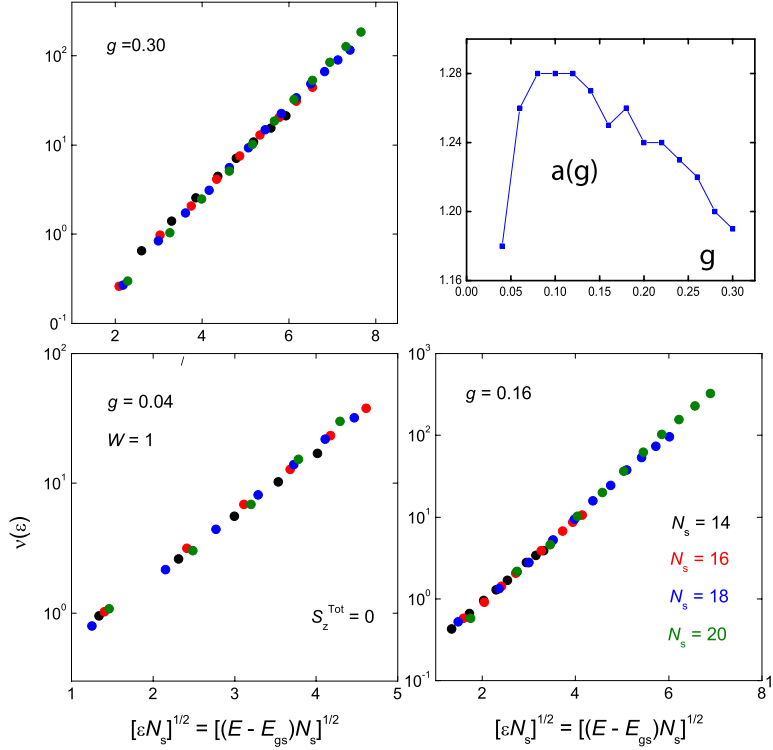}

\caption{Density of states of the Hamiltonian (\ref{H_A}) obtained for $g=0.04$,
$g=0.16$ and $g=0.30$, as indicated on the panels. The coefficient
$a(g)$ is determined by the slope of the straight lines that provide
the best fit to the data points. The upper right panel shows the $a(g)$
dependence. It diplays a maximum for $g$ close to the critical point
$g_{c}$. }

\label{Flo:DOSData} 
\end{figure}

For the quantitative analysis of the numerical results we need the
energy dependence of the total density of states $\nu(E)$ of the
Hamiltonian (\ref{H_A}). As we explain in detail below, at low energies
$E\leq N$ it is given by the formula first found by H. Bethe~\cite{Bethe1936}
for heavy nuclei: 
\begin{equation}
\nu(E)=CN\exp(\alpha\sqrt{EN})\label{eq:nu(E)_0}
\end{equation}
 where $C\sim1$ and $\alpha\sim1$ are functions which depend only
weekly on $g$. This equation and the discussion below assume that
the band width is chosen as $W=1$. The numerical data presented in
Fig. \ref{Flo:DOSData} show that the density of many body levels
of Hamiltonian (\ref{H_A}) are indeed given by Eq.(\ref{eq:nu(E)_0})
for all values of $g$, with a weakly $g$-dependent coefficient $a(g)$.
Because the functional form of $\nu(E)$ is $g$-independent, one
can understand its origin by considering the simple problem of non-interacting
spins, defined by the first term of (\ref{H_A}).

In this approximation the excitation energy is $E=\sum_{i=1,N}\eta_{i}\sigma_{i}$
where $\sigma_{i}$ are equal to 0 or 1, and $\eta_{i}\in[0,1]$ are
quenched random parameters. Let us denote by $\nu(E)$ the typical
density of states. Its Laplace transform , for one given instance
(i.e. one given realization of $\eta_{i}$) is 
\begin{equation}
\tilde{\nu}(s)=\sum_{s_{i}\in\{0,1\}}\exp(-\sum\eta_{i}\sigma_{i}s)\ =\prod_{i}\left(1+e^{-s\eta_{i}}\right)\ .
\end{equation}

For a large system, this density of states self-averages, in the sense
that $\log\nu$ goes to a well defined limit: 
\begin{equation}
\log\tilde{\nu}(s)=N\int_{0}^{1}d\eta\log\left(1+e^{-s\eta}\right)\ .
\end{equation}
 In order to reconstruct the typical density of states using the inverse
Laplace transform, we must find the maximum over $s$ of $Es+N\int_{0}^{1}d\eta\log\left(1+e^{-s\eta}\right)$.The
value of $s$ at which this expression is maximal satisfies the equation
\begin{equation}
\frac{E}{N}=\int_{0}^{1}d\eta\frac{\eta e^{-\tilde{s}\eta}}{1+e^{-\tilde{s}\eta}}
\end{equation}

leading to

\begin{equation}
\nu(E)\approx\frac{1}{\sqrt{E}}\exp\left\{ N_{s}\int_{0}^{1}d\eta\ln\left[1+e^{-\tilde{s}\eta}\right]+E\tilde{s}\right\} \label{eq:nu(E)}
\end{equation}

In the limit of small $E/N$ the maximum is found at large $s$, so
that 
\begin{equation}
\frac{E}{N}\simeq\frac{1}{\tilde{s}^{2}}\int_{0}^{\infty}dx\frac{xe^{-x}}{1+e^{-x}}=\frac{c}{\tilde{s}^{2}}
\end{equation}
 where $c=\frac{\pi^{2}}{12}$. Altogether this shows that the density
of states grows as $e^{\alpha\sqrt{N_{s}E}}$, where 
\begin{equation}
\alpha=2\sqrt{c}=\frac{\pi}{\sqrt{3}}\approx1.8
\end{equation}

\begin{figure}
\includegraphics[width=4in]{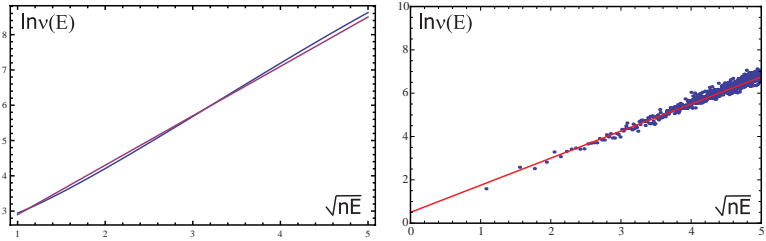}

\caption{Density of states and its approximations. The left panel shows the
density of states given by the full analytical formula (\ref{eq:nu(E)})
and its approximation by the simplified square root dependence $\ln\nu=a\sqrt{NE}+b$
for $N=18$ with $a=1.4$ and $b=1.5$. The right panel shows the
direct numerical simulation of 100 realization for the system of 18
spins in sector $S_{z}=0$ and its fit to a $\ln\nu=a\sqrt{nE}+b$
dependence, with $a=1.25$, $b=0.5$. }

\label{Flo:DOS} 
\end{figure}

For realistic $N$, however, the minimum is somewhat different as
can be seen from the Figure \ref{Flo:DOS} in which we show the full
expression (\ref{eq:nu(E)}) for $N=18$ and its fit to the square
root dependence.

In comparison, the direct numerical simulation of 100 realizations
for $N=18$ system shows a similar behavior with slightly smaller
coefficient $a=1.25-1.35$. The asymptotic (\ref{eq:nu(E)_0}) does
not change if we restrict the states to the sector $S^{z}=0$ computed
numerically, see Figure \ref{Flo:DOS} where we show the results restricted
to this sector.

In the energy range $1\leq E\leq N$, the most probable number $n(E)$
of single-spin excitations whose energies sum up to $E$, is $n(E)\approx\sqrt{EN}$.
Typical single-spin excitation energies are $\epsilon(E)=\sqrt{E/N}$.

At higher energies, density of states reaches it maximum at $E=E_{0}\sim N$,
with $\ln\nu(E_{0})\sim N$, where $E_{0}$ corresponds to the center
of the many-body band. One can relate extensive energy $E=\epsilon N$
of the spin system and its temperature $T$, assuming internal equilibrium:
\begin{equation}
T=\left(\frac{d\ln\nu(E)}{dE}\right)^{-1}=\frac{2}{\alpha}\epsilon(E)\qquad\mathrm{for}\quad E\ll N\label{T_E}
\end{equation}
 Equation (\ref{T_E}) shows that typical single-spin excitation energy
grows with temperature as $\epsilon=\alpha T/2$. Note that $T(E)$
diverges as energy approaches band center, $T^{-1}(E_{0})=0$.

\subsection{Finite-size effect upon the transition line\label{sub:Finite-size-effect-upon}}

\begin{figure}
\includegraphics[width=4in]{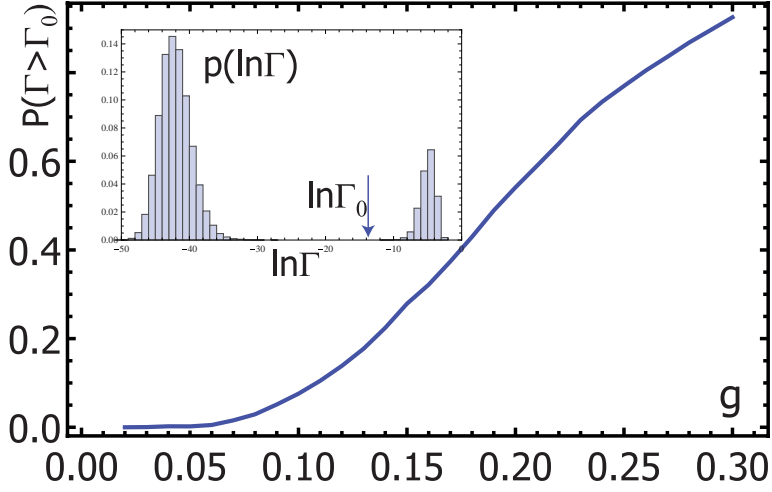}

\caption{ The solution of cavity equations in finite random graphs gives a
widely distributed width $\Gamma$ of single spin flip excitations.
The insert shows the probability distribution of $\Gamma(\omega)$
for $\omega=0.35W$ in a random graph with $N=20$ spins and a coupling
constant $g=0.10$. The right-hand peak in this distribution corresponds
to all graphs in which a significant $\Gamma(\omega)$ is spontaneously
formed. The main panel shows the weight of the graphs in which a significant
$\Gamma(\omega)$ (defined as $\ln\Gamma>\ln\Gamma_{0}$) is formed
as a function of the interaction constant $g$. The resulting value
is only weakly sensitive to the value of the cutoff $\Gamma_{0}$
for most $\omega$ and $g$. This allows us to determine the apparent
critical value of the interaction constant $g(\omega)$ for the finite
systems. The results are shown in Fig.\ref{Flo:PhaseDiagram} as a
dashed line. }

\label{Flo:P(g)} 
\end{figure}

In a finite system the crossover from the Poisson to Wigner-Dyson
statistics is expected to occur when the states become delocalized.
Delocalization implies that the width of the level determined self-consistently
in the cavity approach \cite{Feigelman2010b} is of the order of the
level spacing. A number of finite size effects make the direct comparison
between the analytical results for infinite systems\cite{Feigelman2010b}
and numerical results reported here non-trivial. First, the width,
$\Gamma(E)$, of the level is expected to become exponentially small
as the energy threshold is approached: 
\[
\Gamma(\epsilon)\sim e^{-\frac{\omega_{1}}{\epsilon-\epsilon_{c}(g)}}
\]
 Furthermore, the coefficient,\, $\omega_{1}$ in this equation is
much larger (see Ref.~\cite{Feigelman2010b}) than $\epsilon_{c},$
so the decay rate becomes very small even relatively far from the
transition line. In a finite system with a not-so-small level spacing,
this leads to a very significant shift of the apparent critical energy
due to finite size effects. Second, the width of the levels fluctuate
strongly from one graph realization to another, see insert in Fig.
\ref{Flo:P(g)}. The large peak at small values of $\Gamma$ shows
that in many realizations the level width is essentially zero. The
change in the statistics observed in small systems is due to a relatively
small number of graphs with significant $\Gamma$. This explains why
the crossover is always observed at a value of $r_{n}$ that is close
to the one of Poisson statistics. In order to compute the critical
value of the coupling constant expected in finite systems we evaluated
the probability, $P(\Gamma_{0})$ to find the level width $\Gamma>\Gamma_{0}$
obtained by the solution of cavity equations for the systems of this
size. The result is shown in Fig. \ref{Flo:P(g)} for a typical value
of $\epsilon$ and $\Gamma_{0}.$ The crossing point observed by direct
numerical diagonalization is expected to happen when the probability
$P(\Gamma_{0}=\delta(E))$ becomes non-zero. To avoid the problem
with numerical errors, we used the condition $P(\Gamma_{0}=\delta(E))>\delta_{0}$
with $\delta_{0}=0.004-0.02$ and checked that the results are not
sensitive to the specific value of $\delta_{0}$ in this interval.
Using this condition we take into account both finite size effects
and compute the finite-size corrections to the infinite-system $\epsilon_{c}(g)$
predicted analytically. The final result for the apparent transition
for small sizes is shown in Fig. \ref{Flo:PhaseDiagram} by the green
line, it is shifted with respect to the infinite-size $\epsilon_{c}(g)$
shown by red line by a factor of two.

Finally, when comparing the predictions of the cavity equations with
the results of the exact diagonalization, we have to take into account
that high energy levels of the whole system correspond to many spin
flips: the resulting decay rate is the sum of the decay rates of the
individual flips. Because of the weak (logarithmic) dependence of
the probability $P(\Gamma_{0})$ on $\Gamma_{0}$ this fact has a
very weak effect on the expected transition. Because of the fast (exponential)
dependence of the density of states on the energy, the typical spin
flip contributing to the high energy level has a well defined energy
given by the saddle point solution of section \ref{sub:Density-of-states}:
$\epsilon(E)\sim\sqrt{EW/N}$. This allows us to map the energies,
$E$, studied in the numerical diagonalization to a typical energy
of a single flip, $\epsilon(E)$, and to compare the results of the
numerical diagonalization and solution of cavity equations as shown
in Fig. \ref{Flo:PhaseDiagram}.

\subsection{$S_{z}-S_{z}$ coupling and finite-$T$ phase transition\label{sub:SzSz-coupling-and}}

In the previous papers~\cite{Ioffe2010,Feigelman2010b} we derived
a recursion relation for the intrinsic level widths $\Gamma_{i}$
of single-spin-flip excitations, assuming a local Bethe lattice structure
of the lattice: 
\begin{equation}
\Gamma_{i}=(J^{xy})^{2}\sum_{k(i)}\frac{\Gamma_{k}}{\left(\omega-W|\xi_{k}|\right)^{2}+\Gamma_{k}^{2}}\label{eq:Gamma_Nonlinear_mapping}
\end{equation}
 where $k(i)$ are the $K=Z-1$ neighbours of $i$ in the cavity graph.
The same recursions can be used for a random graph model studied in
the present paper, upon neglecting the contributions of closed loops
(loops are totally absent on the infinite Bethe lattice whereas they
are present with parametrically low density $\sim1/\ln N$ on a random
graph with $N$ sites). The condition of stationarity of the distribution
function $P(\Gamma)$ generated by the mapping (\ref{eq:Gamma_Nonlinear_mapping})
leads to the solution for typical $\Gamma_{typ}(\omega)$ found in
Ref.~\cite{Feigelman2010b}.

In presence of $s_{i}^{z}s_{j}^{z}$ coupling the recursion equations
(\ref{eq:Gamma_Nonlinear_mapping}) should be modified. At nonzero
temperature $T=\beta^{-1}$ it contains the sum over neigboring spin
configurations with their thermal weights: 
\begin{equation}
\Gamma_{i}=(J^{xy})^{2}\sum_{k(i)}\sum_{s_{n(k)}^{z}}\frac{e^{\xi_{n}s_{n(k)}^{z}/T}}{Z_{k}}\frac{\Gamma_{k}}{\left(\omega-\epsilon_{k}\right)^{2}+\Gamma_{k}^{2}}\label{eq:Gamma_Nonlinear_mapping_ZZ}
\end{equation}

where the internal summation over $s_{n(k)}^{z}$ includes all configurations
of spin variables connected to the spin $s_{k}$ by $z-z$ links,
$Z_{k}=\sum_{s_{n(k)}}e^{\beta\xi_{n}s_{n(k)}^{z}}$ are the corresponding
local partition functions, and $\epsilon_{k}=W|\xi_{k}|+J^{z}\sum_{n(k)}s_{n}^{z}$.

The largest contribution to the sum (\ref{eq:Gamma_Nonlinear_mapping})
comes from the site $k$ characterized by $\xi_{k}$ that is closest
to $\omega$. The same holds for the sum (\ref{eq:Gamma_Nonlinear_mapping_ZZ}),
but in addition in this sum the value of $\epsilon_{k}$ varies depending
on the surrounding spins. This increases the probability to find a
resonance. To evaluate the importance of this effect we consider explicitly
the case of $Z=3$ (corresponding to the random graphs diagonalized
in this paper) and the energy at center of the band, $\omega=W/2$.
The critical value of the coupling constant in this case is determined
by the condition $\overline{\ln\Xi}=0$ in the limit of the large
system size. Here 
\begin{equation}
\Xi=\sum_{j\{i\}}\prod_{j}\left(\frac{g}{K}\right)^{2}\frac{1}{Z_{j}}\left\{ \frac{e^{-\beta(\xi_{j1}+\xi_{j2})}}{(\delta\xi_{j}+2\gamma)^{2}}+\frac{2\cosh\beta(\xi_{j1}-\xi_{j2})}{\delta\xi_{j}^{2}}+\frac{e^{\beta(\xi_{j1}+\xi_{j2})}}{(\delta\xi_{j}-2\gamma)^{2}}\right\} \label{Xi}
\end{equation}
 is the relaxation rate induced deep in the system center by the infinitisimally
small couplings at the boundary, $Z_{j}=4\cosh\beta\xi_{j1}\cosh\beta\xi_{j2}$,
$\gamma=J^{z}/W$, $\beta=2T/W$ and $\delta\xi_{j}=\xi_{j}-\omega$.
Performing the same steps as in Ref.~\cite{Feigelman2010b} we average
over the distribution of $\xi$ using the replica method and we get
the condition for the critical value of $g(T)$: 
\begin{eqnarray*}
g & = & K\exp\left[-\frac{1}{2}\min_{x}f(x)\right]\\
f & = & \frac{1}{x}\ln\left\{ K\int d\xi d\xi_{1}d\xi_{2}\left[\frac{1}{Z(\xi_{1},\xi_{2})}\left(\frac{e^{-\beta(\xi_{1}+\xi_{2})}}{(\delta\xi+\gamma)^{2}}+\frac{2\cosh\beta(\xi_{1}-\xi_{2})}{\delta\xi^{2}}+\frac{e^{\beta(\xi_{1}+\xi_{2})}}{(\delta\xi-\gamma)^{2}}\right)\right]^{x}\right\} 
\end{eqnarray*}
 The inversion of the obtained function $g(T)$ provides the $T_{I}(g)$
dependence shown in Fig.\ref{Flo:PhaseDiagramSzSz} by the thick violet
line.

 \bibliographystyle{plain}
\bibliography{ManyBodyDecoherence}

\begin{thebibliography}{10}

\bibitem{Altshuler1997}
Boris~L. Altshuler, Yuval Gefen, Alex Kamenev, and Leonid~S. Levitov.
\newblock Quasiparticle lifetime in a finite system: A nonperturbative
  approach.
\newblock {\em Phys. Rev. Lett.}, 78(14):2803--2806, Apr 1997.

\bibitem{Anderson1958}
P.~W. Anderson.
\newblock Absence of diffusion in certain random lattices.
\newblock {\em Phys. Rev.}, 109(5):1492--1505, Mar 1958.

\bibitem{Anderson1959}
P.W. Anderson.
\newblock Theory of dirty superconductors.
\newblock {\em Journal of Physics and Chemistry of Solids}, 11(1-2):26 -- 30,
  1959.

\bibitem{Basko2006}
D.M. Basko, I.L. Aleiner, and B.L. Altshuler.
\newblock Metal-insulator transition in a weakly interacting many-electron
  system with localized single-particle states.
\newblock {\em Annals of Physics}, 321(5):1126 -- 205, 2006.

\bibitem{Baturina2008a}
T.I. Baturina, A~Yu. Mironov, V.~M. Vinokur, M.~R. Baklanov, and C.~Strunk.
\newblock Hyperactivated resistance in titanium nitride films on the insulating
  side of the disorder-driven superconductor-insulator transition.
\newblock {\em JETP Lett.}, 88:752, 2008.

\bibitem{Bethe1936}
H.~A. Bethe.
\newblock An attempt to calculate the number of energy levels of a heavy
  nucleus.
\newblock {\em Phys. Rev.}, 50:336--, 1936.

\bibitem{Bouadim2011}
Karim Bouadim, Yen~Lee Loh, Mohit Randeria, and Nandini Trivedi.
\newblock Single and two-particle energy gaps across the disorder-driven
  superconductor-insulator transition.
\newblock {\em cond-mat arXiv:1011.3275}, 2010.

\bibitem{Scardiccio2011}
F.~Buccheri, A.~De~Luca, and A.~Scardicchio.
\newblock On the structure of typical states of a disordered richardson model
  and many-body localization.
\newblock {\em arXiv:1103.3431}, Mar 2011.

\bibitem{Canovi2011}
Elena Canovi, Davide Rossini, Rosario Fazio, Giuseppe~E. Santoro, and
  Alessandro Silva.
\newblock Quantum quenches, thermalization, and many-body localization.
\newblock {\em Phys. Rev. B}, 83:094431, Mar 2011.

\bibitem{Fazio2001}
R.~Fazio and H.~van~der Zant.
\newblock Quantum phase transitions and vortex dynamics in superconducting
  networks.
\newblock {\em Physics Reports-review Section of Physics Letters},
  355(4):235--334, December 2001.

\bibitem{Feigelman2010a}
M.~V. Feigel'man, L.~B. Ioffe, V.~E. Kravtsov, and E.~Cuevas.
\newblock Fractal superconductivity near localization threshold.
\newblock {\em Annals of Physics}, 325(7, Sp. Iss. SI):1390--1478, July 2010.

\bibitem{Feigelman2007}
M.~V. Feigel'man, L.~B. Ioffe, V.~E. Kravtsov, and E.~A. Yuzbashyan.
\newblock Eigenfunction fractality and pseudogap state near the
  superconductor-insulator transition.
\newblock {\em Phys. Rev. Lett.}, 98(2):027001, Jan 2007.

\bibitem{Feigelman2010b}
M.~V. Feigel'man, L.~B. Ioffe, and M.~M\'ezard.
\newblock Superconductor-insulator transition and energy localization.
\newblock {\em Phys. Rev. B}, 82(18):184534, Nov 2010.

\bibitem{Fisher1990a}
Matthew P.~A. Fisher.
\newblock Quantum phase transitions in disordered two-dimensional
  superconductors.
\newblock {\em Phys. Rev. Lett.}, 65:923--926, Aug 1990.

\bibitem{Fisher1990b}
Matthew P.~A. Fisher, G.~Grinstein, and S.~M. Girvin.
\newblock Presence of quantum diffusion in two dimensions: Universal resistance
  at the superconductor-insulator transition.
\newblock {\em Phys. Rev. Lett.}, 64:587--590, Jan 1990.

\bibitem{Gantmakher2010}
V.~F. Gantmakher and V.~T. Dolgopolov.
\newblock Superconductor - insulator quantum phase transition.
\newblock {\em Physics - Uspekhi}, 53:3, 2010.

\bibitem{Gornyi2005}
I.~V. Gornyi, A.~D. Mirlin, and D.~G. Polyakov.
\newblock Interacting electrons in disordered wires: Anderson localization and
  low-$t$ transport.
\newblock {\em Phys. Rev. Lett.}, 95(20):206603, Nov 2005.

\bibitem{Ioffe2010}
L.~B. Ioffe and Marc M\'ezard.
\newblock Disorder-driven quantum phase transitions in superconductors and
  magnets.
\newblock {\em Phys. Rev. Lett.}, 105(3):037001, Jul 2010.

\bibitem{Luttinger1964}
J.~M. Luttinger.
\newblock Theory of thermal transport coefficients.
\newblock {\em Phys. Rev.}, 135(6A):A1505--A1514, Sep 1964.

\bibitem{Oganesyan2007}
Vadim Oganesyan and David~A. Huse.
\newblock Localization of interacting fermions at high temperature.
\newblock {\em Phys. Rev. B}, 75(15):155111, Apr 2007.

\bibitem{Ovadia2009}
M.~Ovadia, B.~Sac\'ep\'e, and D.~Shahar.
\newblock Electron-phonon decoupling in disordered insulators.
\newblock {\em Phys. Rev. Lett.}, 102(17):176802, Apr 2009.

\bibitem{Paramanandam2011}
J.~Paramanandam, M.T. Bell, , L.B. Ioffe, and M.E. Gershenson.
\newblock Magnetic field driven quantum transitions in josephson arrays.
\newblock {\em to be published}, 2011.

\bibitem{Sacepe2011b}
B.~Sacepe.
\newblock {\em Private communication}, 2011.

\bibitem{Sacepe2010}
B.~Sacepe, C.~Chapelier, T.I. Baturina, V.M. Vinokur, M.R. Baklanov, and
  M.~Sanquer.
\newblock Pseudogap in a thin film of a conventional superconductor.
\newblock {\em Nature Communications}, 1:140, 2010.

\bibitem{Sacepe2011a}
B.~Sacepe, T.~Dubouchet, C.~Chapelier, M.~Sanquer, M.~Ovadia, D.~Shahar,
  M.~Feigel'man, and L.~Ioffe.
\newblock Localization of preformed cooper pairs in disordered superconductors.
\newblock {\em Nature Physics}, 7(3):239 -- 44, 2011.

\bibitem{Sambandamurthy2005}
G.~Sambandamurthy, L.~W. Engel, A.~Johansson, E.~Peled, and D.~Shahar.
\newblock Experimental evidence for a collective insulating state in
  two-dimensional superconductors.
\newblock {\em Phys. Rev. Lett.}, 94(1):017003, Jan 2005.

\bibitem{Sambandamurthy2004}
G.~Sambandamurthy, L.~W. Engel, A.~Johansson, and D.~Shahar.
\newblock Superconductivity-related insulating behavior.
\newblock {\em Phys. Rev. Lett.}, 92(10):107005, Mar 2004.

\bibitem{Shahar1992}
D.~Shahar and Z.~Ovadyahu.
\newblock Superconductivity near the mobility edge.
\newblock {\em Phys. Rev. B}, 46(17):10917--10922, Nov 1992.

\end{thebibliography}

\end{document}